\documentclass{emulateapj}

\shorttitle{Turbulence production by relativistic ion beams}
\shortauthors{Niemiec et al.}
\usepackage{natbib}
\newcommand{\mkp}{ }
\newcommand{\jn}{ }
\newcommand{\jnn}{ }

\begin{document}

\title{Aperiodic Magnetic Turbulence Produced by Relativistic Ion Beams}

\author{Jacek Niemiec}
\affil{Instytut Fizyki J\c{a}drowej PAN, ul. Radzikowskiego 152,
 31-342 Krak\'{o}w, Poland}
\email{Jacek.Niemiec@ifj.edu.pl}
\author{Martin Pohl}
\affil{Department of Physics and Astronomy, Iowa State University, Ames, IA
50011}
\author{Antoine Bret}
\affil{ETSI Industriales, Universidad de Castilla - La Mancha,
13071 Ciudad Real, Spain}
\and
\author{Thomas Stroman}
\affil{Department of Physics and Astronomy, Iowa State University, Ames,
IA 50011}

\begin{abstract}
 Magnetic-field generation by a relativistic ion beam propagating through an
electron-ion plasma along a homogeneous magnetic field is investigated with 2.5D
high-resolution particle-in-cell (PIC) simulations. The studies test 
predictions of a strong amplification of short-wavelength modes of magnetic
turbulence upstream of nonrelativistic and relativistic parallel shocks
associated with supernova remnants, jets of active galactic nuclei, and gamma-ray bursts.
We find good agreement in {\mkp the} properties of the turbulence 
observed in our simulations compared with the dispersion relation calculated
for linear waves with arbitrary orientation of ${\vec k}$. {\mkp Depending on the
parameters,} the backreaction on the ion beam leads to
filamentation of the ambient plasma and the beam, which in turn influences the
properties of the magnetic turbulence. {\jn 
For mildly- and ultra-relativistic beams, the instability saturates at field amplitudes
a few times larger than the homogeneous magnetic field strength.} This result matches our
recent studies of nonrelativistically drifting, hot cosmic-ray particles upstream of
supernova-remnant shocks which indicated only a moderate magnetic-field
amplification by nonresonant instabilities. We also demonstrate that the aperiodic
turbulence generated by the beam can provide efficient particle scattering with a rate
compatible with Bohm diffusion. Representing
the ion beam as a constant external current, i.e. excluding a backreaction of
the magnetic turbulence on the beam, we observe non-resonant parallel modes with
wavelength and growth rate as predicted by analytic
calculations. In this unrealistic setup the magnetic field is amplified to
amplitudes far exceeding the homogeneous field, {\mkp as observed in} recent MHD and
PIC simulations. 
\end{abstract}


\section{INTRODUCTION}
Collisionless shocks are the acceleration sites of energetic particles
responsible for high-energy emission of astrophysical objects and contributing
to the flux of cosmic rays detected at Earth. 
Nonthermal particle populations at nonrelativistic shocks of supernova remnants
(SNRs) are believed to 
be generated {by diffusive shock acceleration (DSA). The 
particle spectra thus produced indeed} agree well with those deduced
 from radio--to--X-ray electron synchrotron emission of SNRs
\citep{reynolds08}. The
high efficiency of the DSA mechanism together with considerations of the global
energetics of a supernova explosion make the forward shocks of shell-type SNRs
the prime candidates for the sources of Galactic cosmic rays (CRs). Nonthermal
power-law particle spectra attributed to relativistic electrons are also
inferred from modeling the electromagnetic emission from astrophysical sources
harboring relativistic shock waves, such as jets of active galactic nuclei
(AGN) and gamma-ray bursts (GRBs) \citep{meszaros02}. 

Efficient particle acceleration at shocks invariably requires a continuous
excitation of magnetic 
turbulence in the upstream region, which serves as a scattering medium to
confine the energetic 
particles to the shock region for further acceleration \citep{md01}. Turbulent magnetic
fields of amplitude much larger than the homogeneous interstellar field are
needed upstream of SNR shocks to account for protons of energies up to
{and beyond} the
"knee" at 10$^{15}$ eV in the cosmic-ray spectrum. Recent X-ray observations of
several young SNRs give evidence that indeed highly amplified fields exist
downstream of SNR forward shocks  (see \citet{reynolds08}, and references therein). 
Downstream of ultrarelativistic GRB external shocks the magnetic fields must
also be amplified orders of magnitude beyond shock-compression levels to explain
GRB afterglow spectra and light curves. Even in the preshock medium magnetic
fields of milligauss strengths are required to account for the observed X-ray
afterglows \citep{li06}. 

A plausible scenario for magnetic-field generation assumes that cosmic ray
particles accelerated at the shock drift as an ensemble relative to the upstream
plasma and trigger a variety of instabilities that may lead to the growth of a
turbulent field component \citep{bl01,bell04}. The distribution function of the 
cosmic rays is shaped by the scattering rate in the self-excited field upstream,
thus forcing a nonlinear 
relationship between the upstream plasma, the energetic particles, and
small-scale 
electromagnetic fields. The upstream field would subsequently be advected and
compressed downstream of the shock and possibly further amplified by fluid
instabilities in the downstream plasma 
\citep{gia07,zir08,couch08}. While a full modeling of the upstream region is
elusive to date,
simulations of turbulence build-up using prescribed distribution functions for
the
upstream plasma and the cosmic rays can be invaluable tools for the study of the
saturation processes and levels, as well as the backreaction of the evolved
turbulence on the 
particles. The system in which {a} population of cosmic rays slowly drifts
relative to the 
upstream plasma has been studied with MHD simulations, {\jn which represented cosmic rays
with a constant external current}
\citep{bell04,bell05,zira08,rev08}
and with first-principles particle-in-cell (PIC) simulations assuming a constant cosmic-ray current \citep{ohira09} or including the full dynamics of the energetic particles
\citep{niem08,stroman,riqu09}. They confirmed the quasi-linear predictions by
\citet{bell04}, who showed that for the parameters of young SNRs, magnetic
turbulence would appear in a form of nonresonant, circularly polarized, and
aperiodic transverse waves. Numerical simulations analyzing the nonlinear
evolution of the system found that the turbulence growth eventually saturates, but
the exact saturation levels differ between the approaches, {the full PIC 
simulations typically yielding considerably lower field amplitudes than MHD studies with 
constant CR currents}. 

Here we report kinetic (PIC) simulations of the interaction
between 
the far-upstream plasma and a cold dilute relativistic beam of particles
streaming along a homogeneous background magnetic field. We assume that the beam
is composed of cosmic-ray ions, and the current {\mkp and charge} carried by the beam is balanced
by electrons of the background medium. The situation is
relevant to the upstream region of both relativistic and nonrelativistic shocks
undergoing efficient particle acceleration. In a SNR shock environment, it
applies to the most energetic cosmic rays accelerated at the shock which stream
far upstream of the free-escape boundary. In this case, a predominantly ionic
cosmic-ray component results from the character of injection processes at
nonrelativistic shocks. Upstream of relativistic shocks of GRBs and AGN,
distributions of particles accelerated in a wide energy range are highly
anisotropic. This is because particles and the shock move close to the speed of
light, and a deflection of a particle trajectory by an angle greater than 
$\Gamma_{sh}^{-1}$ allows the shock to overtake the particle. 
Therefore, in the upstream rest frame the nonthermal particles are highly beamed and their transverse momenta are a factor of $\Gamma_{sh}^{-1}$  smaller than the momenta along the shock direction. {\jnn We approximate this situation by assuming that the CR beam is cold. This assumption better holds for the freshly accelerated particles in the far-upstream region, whose transverse momenta are much smaller than $\Gamma_{sh}^{-1}$  times the parallel momentum; the highest-energy particles escape from the precursor and therefore have an anisotropic distribution in the shock rest frame, which is further enhanced by the shock curvature.} Some instabilities, e.g. filamentation, depend sensitively on the transverse temperature of the CR beam, and care must be exercised in extrapolating our simulation results to situations in which the CR-beam properties are somewhere between those of the cold beam studied here and the very hot, but slow beam investigated earlier \citep{niem08,stroman,riqu09}. {\jnn However,} PIC simulations of relativistic shocks in electron-ion plasmas suggest that filamentation indeed occurs only far upstream of the shock {\jnn(Spitkovsky 2008a; 2008b; see also Medvedev \& Zakutnyaya 2009), and it is generated by a warm ($p_\perp \lesssim p_\parallel/\Gamma_{sh}$) ion beam. We may therefore expect that our assumption of a cold CR beam remains a valid approximation for systems with warm CR beams.} 
Furthermore, our setup applies to the cosmic-ray ions whose energies are larger
than the upper limit on the energy of electrons accelerated by the shock, which
is imposed by {radiative energy losses} {\mkp \citep[e.g.][]{li06}}.
The highly energetic ions will thus reach
farther upstream than the CR electrons and the return current will be provided
by the ambient electrons.

Note that the applicability of the system under study to relativistic
astrophysical sources relies on the ability of relativistic shocks to accelerate
particles to very high energies. Although the first-order Fermi process at such
shocks is widely considered to be the source of cosmic rays, recent studies in
the test particle approximation \citep{nie06a,nie06b,lem06} and using
PIC simulations \citep{sir09} show that this
mechanism can operate only in quasi-parallel or weakly magnetized shocks. If the
GRB or AGN outflows are strongly magnetized{\jn /quasi-perpendicular}, some other processes must be
responsible for particle acceleration (e.g., magnetic reconnection), and our
results do not apply.  

It is known from studying non-relativistic beams in interplanetary 
space that a competition arises between resonant and nonresonant modes, which exert different backreactions on the beam
\citep{winske}. For the
case of a monoenergetic, unidirectional distribution of streaming cosmic rays,
the rates for {the resonant growth of} Alfv\'{e}nic \citep{ps2000} and
electrostatic \citep{pls02} turbulence
have been derived using quasilinear theory. Based on an analytical treatment,
\citet{rev06}
found that also in this case nonresonant, purely growing modes may be expected
to be significantly
faster, although the growth rate falls off with the temperature of the
background medium.
Application of this mechanism to the external GRB shocks was phenomenologically
studied by \citet{mil06}, who concluded that CR-driven turbulence may account
for the levels of amplified magnetic fields inferred from these sources. 
We have performed a series of two-dimensional simulations for this setup to
explore the relationship between this instability and that found for drifting cosmic rays,
and to determine {\jn the mutual  backreaction between the magnetic turbulence and the CR beam.
The interaction of a cold relativistic ion beam {\mkp is studied} in the limit
of a magnetized background plasma, for which the results of the analytical calculations of
\citet{rev06} apply.}

{\jn The simulation setup is described \S 2 and the results of the linear kinetic analysis of the system are presented in \S 3.  In \S 4 the simulation results are presented. The differences in the properties of the system between the runs representing the CR beam with a constant external current and the fully kinetic simulations are discussed in \S 4.1 and \S 4.2. The detailed properties of the magnetic turbulence and the evolution of particle phase-space distributions are then presented in \S 4.3 based on the results for mildly relativistic ion beams.  We conclude with a summary and
discussion in \S 5.}

\section{SIMULATION SETUP}
The code used in this study is a 2.5D (2D3V) version of the relativistic
electromagnetic particle code TRISTAN with MPI-based parallelization
\citep{buneman93,niem08}.  
In the simulations a cold, relativistic, and monoenergetic cosmic-ray ion beam
with Lorentz 
factor $\gamma_{CR}$ {\jn (velocity $v_{CR}$)}
and number density $N_{CR}$ streams along a homogeneous magnetic field
$B_{\parallel 0}$ 
relative to the ambient electron-ion plasma. The ions of the ambient medium have
a thermal distribution with number density $N_i$, in thermal equilibrium with
the electrons. The electron 
population with density $N_e=N_i+N_{CR}$ contains the excess electrons required
to provide charge-neutrality and drifts with $v_d=v_{CR}N_{CR}/N_e$ with
respect to the background ions, so it provides a return current balancing the
current carried by
the ion beam. 
We have explored the system in the limit of a magnetized
background plasma, $\omega\ll \Omega_i$ (see \citep{rev06}). Specifically, we assumed 
$\gamma_{max}/\Omega_i = 0.2$, where 
\begin{equation} 
\label{e1}
\gamma_{max}=\Im\omega\approx {1\over 2}\,{{v_{CR}\,N_{CR}}\over {v_A\, N_i}}\,\Omega_i
\end{equation}
is the growth rate of the most unstable nonresonant mode, $\Omega_i$ is the ion
gyrofrequency, and
$v_{A}=[B_{\parallel 0}^2/\mu_0 (N_em_e+N_im_i)]^{1/2}$ is the plasma Alfv\'{e}n
velocity. The relativistic cosmic-ray populations represent 
very dilute ion beams which we study with density ratios $N_i/N_{CR}=50$ and $125$,
and Alfv\'{e}n velocities
$v_A = c/20$ and $v_A = c/50$, respectively (for which the ratio 
$\omega_{pe}/\Omega_e=4.4$ and $11.0$, respectively). 
The simulations have been performed for the case of an ultrarelativistic beam
with $\gamma_{CR}=300$ and a slower beam with $\gamma_{CR}=20$. We have also
studied the case in which the beam is represented by
a constant uniform external current, so that {the} backreaction of the magnetic
turbulence on the cosmic-ray beam is suppressed. {The parameters of all simulation
runs described here} are summarized in Table 1. 

To ensure the numerical accuracy of our simulations, we use a total of 16
particles per cell, and we apply the splitting method for the beam particles.
The density ratio between {simulated cosmic-ray and ambient particles
on the grid is $1/3$, and
weights are applied to each beam particle to match the desired $N_{CR}/N_{i}$.
The same weights are
used for the excess electrons $\delta N_e=N_e-N_i=N_{CR}$, and thus
each ion particle can be initialized at the same location as the
corresponding electron}  for the identically zero initial charge
density.
The electron skindepth $\lambda_{se}=c/\omega_{pe}=4\Delta$, where 
$\omega_{pe}=(N_e e^2/m_e\epsilon_0)^{1/2}$ is the electron plasma frequency
and $\Delta$ is the grid cell size. We further assume a reduced ion-electron
mass ratio 
$m_i/m_e=20$. This choice allows us to clearly separate the plasma and
turbulence scales and yet use a computational box that can contain several
wavelengths of the most unstable mode 
\begin{equation} 
\label{e2}
\lambda_{max}\approx 2\pi(\gamma_{max}/\Omega_i)^{-1}\lambda_{si}, 
\end{equation}
where $\lambda_{si}$ is the ion skindepth. 
As {the} results of our linear analysis of \S 3 show, the one-dimensional dispersion
relations obtained 
by \citet{rev06} capture the physical properties of the system,
provided the Lorentz factor of the ion beam is {very} large or 
a constant external current is applied. In this study we perform
simulations for an ultrarelativistic beam and for the case of a constant
external cosmic-ray current to cross-check those results as well as the validity
of our approach. For these runs we use smaller computational grids: ($L_{\rm
x}, L_{\rm y}) = (7.4 \lambda_{max}, 5.7\lambda_{max})$ for runs A and D, and
($L_{\rm x}, L_{\rm y}) = (4.6 \lambda_{max}, 3.3\lambda_{max})$ for run B.
Larger grids with ($L_{\rm x}, L_{\rm y}) = (10.2 \lambda_{max},
7.4\lambda_{max})$ are used to investigate the systems with mildly relativistic
cosmic-ray beams, which have not been studied with PIC simulations before.
Cosmic-ray beam ions move in the $-x$-direction, antiparallel to the
homogeneous 
magnetic field $B_{\parallel 0}$.
Periodic boundary conditions are assumed for all boundaries. In all simulations
the time step {is} $\delta t=0.1125/\omega_{pe}$, and the inverse maximum growth rate
of the nonresonant modes $\gamma_{max}^{-1}=3975\delta t$ for runs A--C and
$9938\delta t$ for runs D and E.

Our previous work on similar systems as well as additional test runs performed
for the current study ensures that our results are not affected by a particular
choice of simulation parameters, e.g., electron-ion mass ratio, number of
particles per cell, or the electron skindepth. The validity of two-dimensional
simulations in capturing the essential physics has been demonstrated in
\citet{niem08}.

\section{LINEAR ANALYSIS \label{lin}}
The growth rate and wavelength of the most unstable purely-growing
nonresonant mode given by 
equations \ref{e1} and \ref{e2} were obtained by \citet{bell04} and \citet{rev06}
using linear kinetic analysis in
the limit of a cold ambient plasma and only for wavevectors $k_\parallel$ parallel to
$B_{\parallel 0}$. We have numerically calculated the growth rates for
arbitrary 
orientation of the wavevector, $\vec{k}$, in the zero-temperature limit {\jn (for details of the calculations see,
e. g., \citep{bret09})}. 
In Figure 1 we show the growth rates in the reduced wavevector space
$(Z_\parallel, Z_\perp)$, 
$Z_i=k_iv_{CR}/\omega_{pe}$, that is contained in our simulation box, for
beams moving along a homogeneous magnetic field of strength given by the
Alfv\'{e}n velocity of $v_A = c/20$.   
In each case
the dominant unstable mode
is the electrostatic Buneman mode between background ions and drifting
electrons 
($Z_\parallel\gtrsim 2$). The growth rate
of this mode is about $10^2$ times larger, and its wavelength about $1.25\times
10^3$ shorter than that of the nonresonant mode {(see Eqs. \ref{e1}
and \ref{e2})}. Our simulations do not fully resolve this
mode, because the wavelength of maximum growth corresponds to half a cell
($Z_\parallel\approx 55$) on our computational grid,
and we are able to see longer wavelength modes only with somewhat smaller growth
rates. However, the Buneman instability is very sensitive to thermal effects and
should
saturate if the thermal velocity of ambient particles becomes comparable to
their relative
drift velocity. The initial electron thermal velocity in the simulations is thus
set
to values $v_{e,th}\lesssim v_d$ to ensure the quick saturation and dissipation
of this unstable
mode. Note that such plasma parameters well reproduce the real conditions in
astrophysical
objects, and the Buneman mode will be relevant {only if the beam density is
high, because then $v_d$ will be high as well.}

\begin{figure}
\plotone{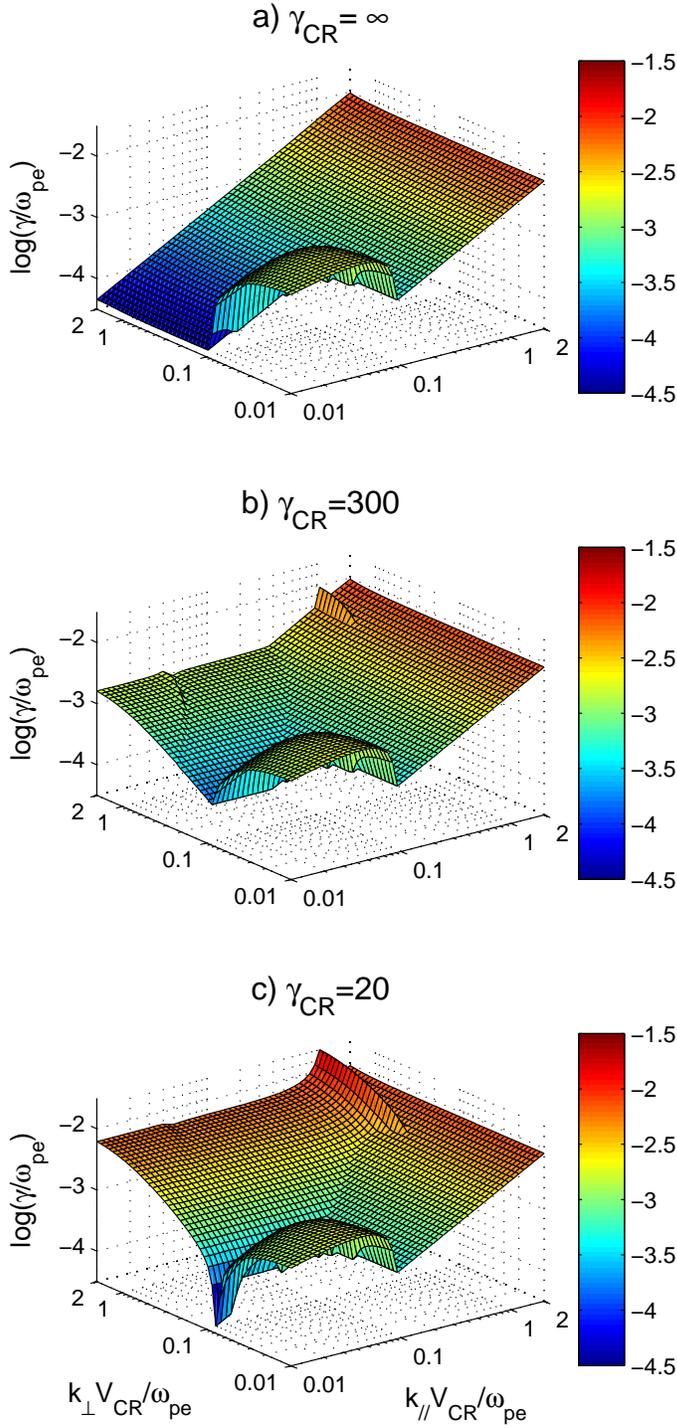}
\caption{{Linear growth rate $\gamma$ in units of the electron plasma frequency, 
$\omega_{pe}$, as function of the flow-aligned and perpendicular 
($k_\parallel$ and $k_\perp$) wavevectors for the parameters used in simulations
A, B, and C (see Table~1).} {\jn For a given $\vec{k}$, the growth rate of the most unstable mode is plotted. The figures} show the
modes whose wavelengths are well contained in the simulation box.
\label{fig1}}
\end{figure}

The nonresonant mode, {\mkp which we are chiefly interested in,} is visible at $Z_\perp <
0.1$ and shows
a broad peak centered at $Z_\parallel\approx 0.05$, corresponding to the
estimate
given by Eq.~\ref{e2}. However, {if the CR beams are treated fully
kinetically} ($\gamma_{CR}=300$ and $20$ -- Fig. \ref{fig1}b and
Fig. \ref{fig1}c, respectively), the nonresonant mode is not dominant even in the
limited wavevector space {covered in our simulations.}
In fact, the strongest growth occurs for $0.3 < Z_\perp < 1$, almost
independent of $Z_\parallel$. 
The very peaked growth at $Z_\parallel\approx 1$ pertains to the Buneman 
instability between relativistic ion beam and ambient electrons. The growth at
smaller 
$Z_\parallel$ represents the filamentation of the ambient plasma and the ion
beam. 
The appearance of these fast-growing modes modifies the system, and one
should expect that the properties of the nonresonant mode emerging in the
nonlinear stage
differ from those predicted in the analytical calculations {\jn by \citet{rev06}}. This is in fact what
is observed in our simulations. {\mkp It should be noted, though, that for warm CR beams
filamentation {\jnn might be} suppressed, and therefore may not play a role in the precursor region
close to the shock.}

Note that the growth rates as shown in Figure 1 depend on the parameters of the
system
under study. In particular, for ultrarelativistic beams (Fig. \ref{fig1}b) the
growth rate for the filamentation 
modes is much smaller than for $\gamma_{CR}=20$.
Hence, for $\gamma_{CR}\gtrsim 100$ we primarily have a competition between the
nonresonant mode
and the Buneman modes. If we replace
the ion beam by an constant external current (which somewhat corresponds
to $\gamma \gg 1$, Fig. \ref{fig1}a), then the ambient electrons do not interact
with the ion beam, and a
Buneman instability is not excited. The evolution of the system is then
artificially dominated by the nonresonant instability \citep{ohira09}.  

\section{SIMULATION RESULTS}
\subsection{Simulations with Constant Cosmic-Ray Current}
\begin{figure}
\plottwo{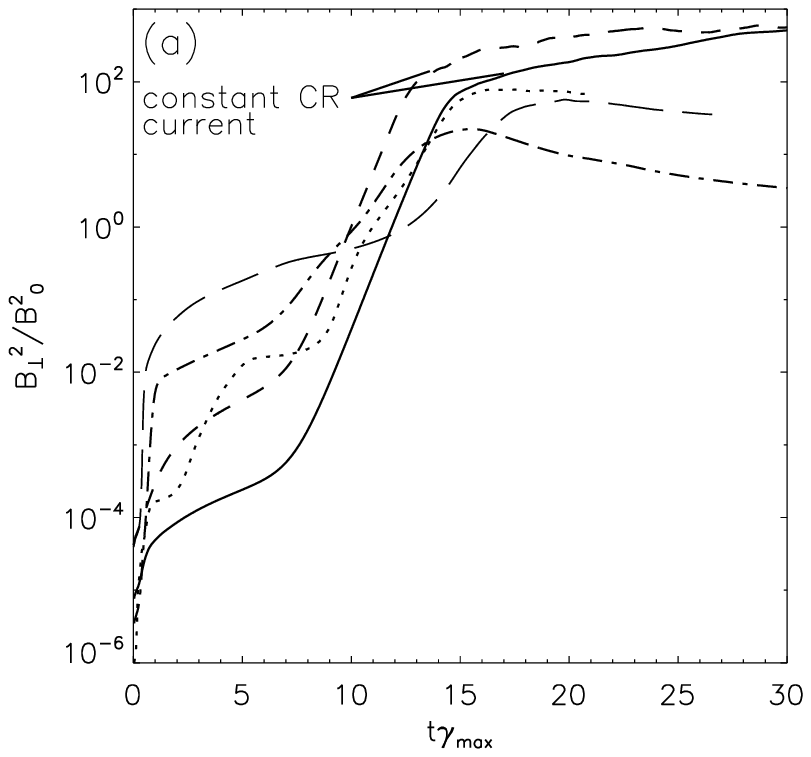}{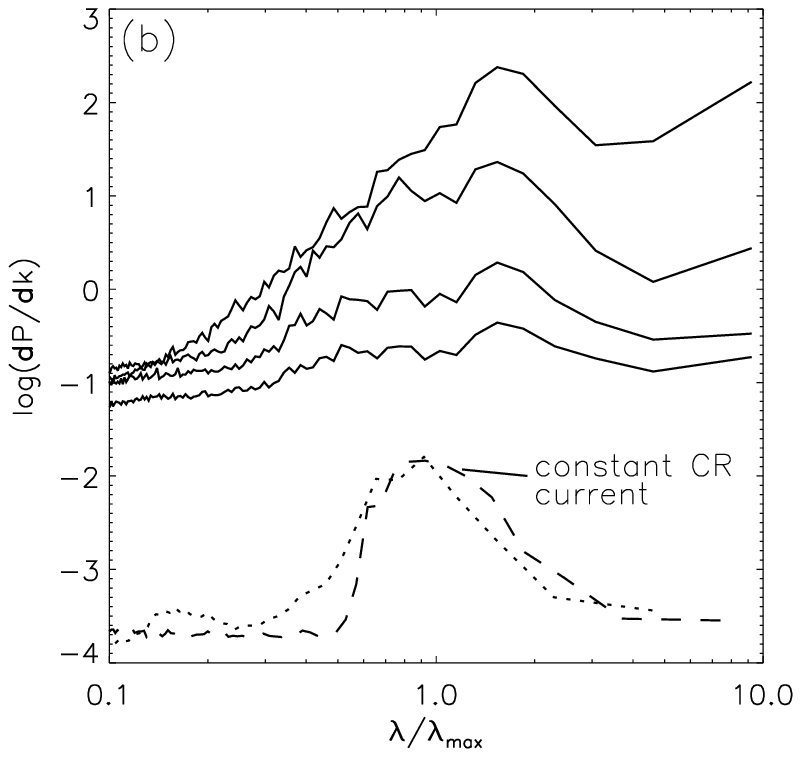}
\caption{Temporal evolution of the energy density in the transverse
magnetic
field component, normalized to the homogeneous field strength (a). 
Time is dimensionless in units of the theoretically
predicted inverse growth rate of the nonresonant mode, $\gamma_{max}^{-1}$.
Simulations with constant
cosmic-ray current are presented with the solid line for a plasma Alfv\'{e}n
velocity $v_A = c/20$ (run A),
 and the dashed line for $v_A = c/50$ (run D). The effects of cosmic-ray
backreaction are shown for the
case of $v_A = c/20$ and the beam Lorentz factor $\gamma_{CR} = 300$ (run B;
dotted) and 
$\gamma_{CR} = 20$ (run C; dashdotted), and for $v_A = c/50$ and $\gamma_{CR} =
20$ (run E; long dashed).
Fourier spectra of the perpendicular magnetic-field component, $B_z$, in
wavelengths along the beam direction for the case with $v_A = c/20$ (runs A-C) (b).
The parallel mode is
expected at the wavelength $\lambda_{max}$. The dashed line shows results for a
constant cosmic-ray
current (run A), and the dotted line for a faster beam with $\gamma_{CR} = 300$
(run B; spectrum shifted for presentation clarity) at the initial stage of the
turbulence growth, $t\gamma_{max}$ = 7 and 9, respectively.
Note the single peak structure centered on $\lambda_{max}$. The solid lines show
the temporal evolution of the spectra for $\gamma_{CR} = 20$ (run C)
during the linear stage ($t\gamma_{max}$ = 6, 7, 9, 11, from bottom to top). The
magnetic turbulence appears in a broad range of wavelengths.
   \label{double_fig}}
\end{figure}

The temporal evolution of the energy density in the transverse magnetic-field
component is shown in Figure~\ref{double_fig}a. 
If the backreaction on the cosmic rays is suppressed, i.e., a constant uniform
external 
current is applied, then a purely-growing parallel mode of magnetic turbulence
appears 
in the plasma. Its growth rate ($\gamma\approx 0.8 \gamma_{max}$ for the two
cases with $v_A = c/20$
and $c/50$, run A and D, respectively) and wavelength (dashed line in Fig.~\ref{double_fig}b)
agree well with those predicted by quasi-linear analytical calculations. 
The mode represents a purely magnetic, circularly polarized, and aperiodic
transverse wave. {The interactions of the magnetic turbulence with the plasma are 
predominantly} related to the
return current carried by the ambient electrons, $\vec{j}_{ret}$. 
The $\vec{j}_{ret}\times\delta B_\perp$ force induces
motions and turbulence in the background plasma, 
{which in the later stages cause} the turbulence
to turn nearly isotropic and highly nonlinear.       
As in the case of drifting cosmic rays \citep{niem08} and nonrelativistic beams
\citep{winske},
the saturation of the magnetic-field growth proceeds via bulk acceleration and
occurs when
the bulk velocity of the background plasma approaches the cosmic-ray ion beam
speed {\jn (see \S 4.3.2.).} {\mkp Note that current and charge balance is still observed
if the cosmic-ray current is chosen constant, and the background plasma is charged by
adding extra electrons to compensate for the charge of the cosmic rays. The plasma thus
''knows'' the cosmic-ray drift speed as that at which the plasma electrons no longer
stream relative to the plasma ions to carry the return current.} The {\jn nonlinear}
amplitude of the field perturbations is {\jn slightly} larger for smaller Alfv\'{e}n velocity,
in agreement with \cite{riqu09}. {\jn However, the magnetic-field amplitudes become comparable at the end of both runs, and reach} 
$\delta B_\perp/B_{\parallel 0}\simeq 25$, which is close to the maximum
obtained with MHD
simulations \citep{bell04,bell05,zira08} {and other PIC simulations \citep{ohira09}, 
in which the cosmic rays were}
also represented by a constant current.

We will now describe the behavior of the system including the 
response of the relativistic cosmic-ray ion beam.

\subsection{Fully Kinetic Simulations}
If the cosmic rays are treated fully kinetically, the dynamics of the system
changes. The
interaction of the ion beam with the plasma quickly leads to plasma and beam
filamentation
which is modified by a Buneman instability between the ion beam and plasma
electrons. 
The Buneman beam-electron interactions produce mainly electrostatic, slightly
oblique turbulence whose wavelength {\jn parallel to the direction of the
beam is in very good agreement with the predictions of our linear analysis, which gives 
 $\lambda=2\pi(v_{CR}/c)\lambda_{se}\sim 25\Delta$ (\S 3)}. 
The mode grows very fast, causing density fluctuations in the beam and electron
plasma. However, in the simulations its amplitude quickly saturates and is subsequently
kept at a moderate level. {\jn These features are in agreement with the known properties of the Buneman modes  (see, e.g., \citet{dieck07} for a detailed discussion of the nonlinear evolution and saturation mechanism of the Buneman instability).}
The Buneman mode dissipates only after filamentation and nonresonant
modes have strongly 
backreacted on the ion beam in the nonlinear stage. 

\subsubsection{Ultrarelativistic Beams}  
The properties of the magnetic turbulence depend on the Lorentz factor of the
beam. 
For an ultrarelativistic beam with $\gamma_{CR} = 300$ (run B; dotted lines in Fig. 2),
the 
filamentation is weak 
and the parallel nonresonant mode appears with the theoretically predicted
wavelength.
Its growth rate is initially $\gamma\approx 0.94 \gamma_{max}$ and decreases
during the
nonlinear evolution. As one can see in Figure 2a, the peak amplitude
of the magnetic-field 
perturbations, $\delta B_\perp/B_{\parallel 0}\simeq 9$, is close to that
obtained with
constant external current (run A; solid line) at the onset of the saturation of the
turbulence growth 
($t\sim 15\gamma_{max}^{-1}$). It appears that in this phase the high beam
Lorentz factor 
provides sufficient stiffness to the ion beam that its backreaction is
suppressed, rendering the
system response similar to that for a constant external current. The similarity
ends when the saturation kicks in, though.
The subsequent dissipation of the turbulence in the run with 
$\gamma_{CR} = 300$ is much stronger than in the case of a constant external
current,
which places in doubt the accuracy of simulations that use a constant external
current 
to describe the highly nonlinear phases in the evolution of the system.

\subsubsection{Mildly Relativistic Beams} 
Results for a system with a mildly relativistic beam with 
$\gamma_{CR} = 20$, and for $v_A = c/20$ (run C) and $c/50$ (run E), are presented in Figure 2a 
with dash-dotted and long-dashed lines, respectively. As our linear analysis of \S 3
shows, the filamentation modes at perpendicular wavevectors $k_\perp \approx 1/\lambda_{se}$
are strong in this case. They cause
filamentation in the ambient plasma and the ion beam, before the nonresonant
parallel modes have 
emerged. As one can see in Figure 2, these modes do not lead to magnetic-field
perturbations of significant amplitude. Nevertheless, their action on the
ambient plasma changes its properties, which considerably influences the
characteristics of 
the purely-growing parallel modes. The nonresonant modes appear in a broad range
of wavelengths around $\lambda_{max}$ (Fig. 2b),
and the growth rate of the magnetic-field perturbations is only $\sim
0.4\gamma_{max}$. 
The backreaction of the turbulence on the system further enhances the
filamentation in the beam 
and the plasma, and leads to the saturation
and dissipation of the magnetic turbulence at a level {\jn a few times the
homogeneous magnetic field strength.}
The peak amplitudes for the two cases with $v_A = c/20$ and $c/50$ are
$\delta B_\perp/B_{\parallel 0}\simeq 4.7$ and
$\delta B_\perp/B_{\parallel 0}\simeq 7.5$, respectively, showing 
that instabilities 
operating in a less-magnetized medium provide a stronger field amplification.
{\mkp It is unclear whether the modification of the parallel mode arises specifically from
filamentation or from any type of perpendicular, small-scale density fluctuations,
including preexisting turbulence. We can therefore not reliably predict the behavior of
a system containing a warm cosmic-ray beam, for a example the denser parts of a
cosmic-ray precursor to an astrophysical shock.}

\subsection{Aperiodic Magnetic Turbulence Produced by Mildly Relativistic
Beams} 
\subsubsection{Spectral Properties of the Turbulence}
\begin{figure}
\plotone{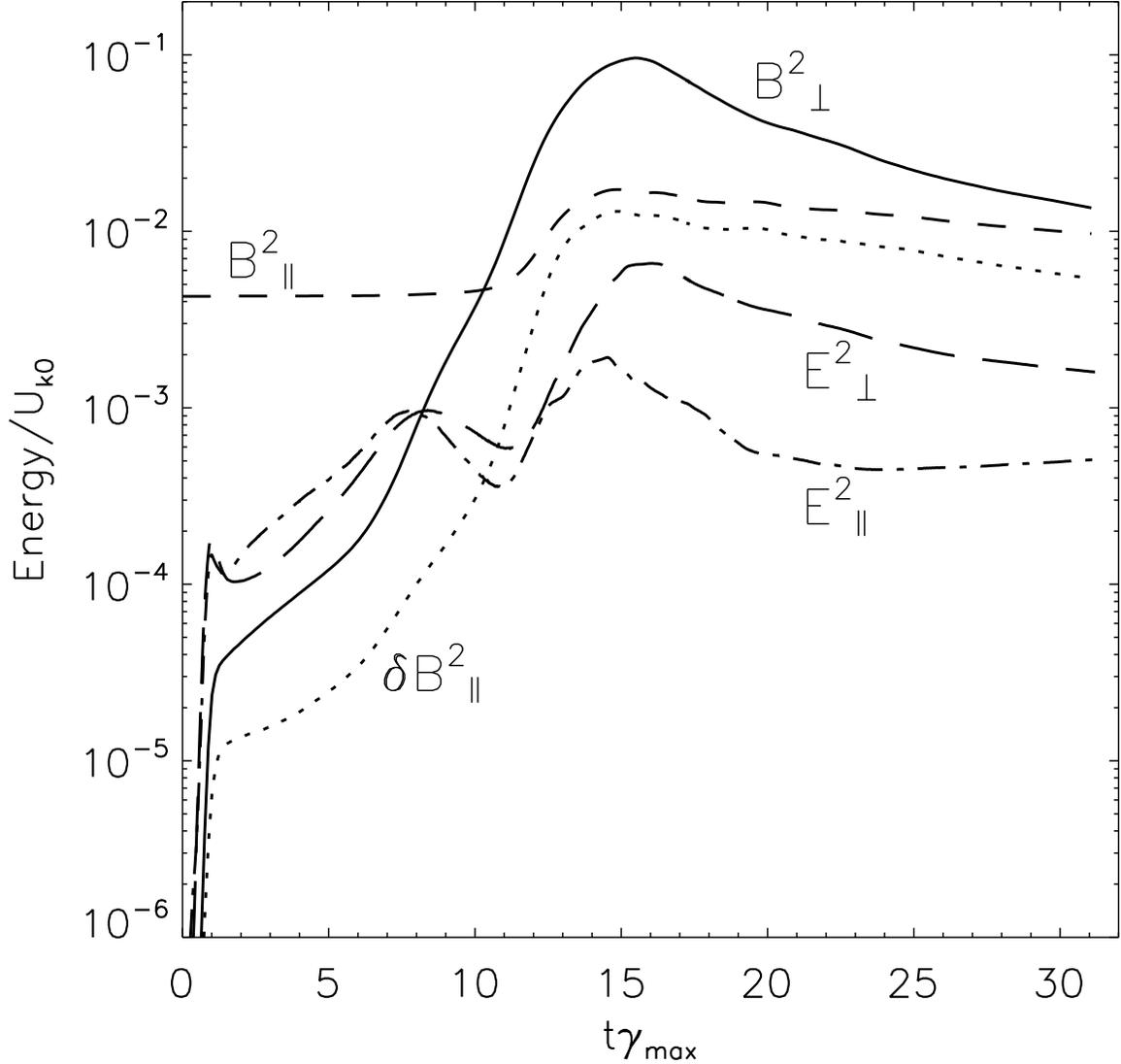}
\caption{Temporal evolution 
of the average energy density in electromagnetic fields, normalized to the 
initial total kinetic energy in the system, $U_{k0}$, for the run with
$\gamma_{CR} = 20$ and $v_A = c/20$ (run C; compare Fig. \ref{double_fig}b).
Here
$B^2_{\perp}$ and  $E^2_{\perp}$ indicate the magnetic and electric energy
densities in the transverse components, i.e.,
$\langle B_y^2+B_z^2\rangle/(2\mu_0)$ and correspondingly for the electric
field.
The quantities $B^2_{\parallel }$, $\delta B^2_{\parallel}$, and 
$E^2_{\parallel}$ are defined analogously, {but
$\delta B^2_{\parallel}$ refers to the turbulent component only.} 
\label{en16}}
\end{figure}

\begin{figure}
\plotone{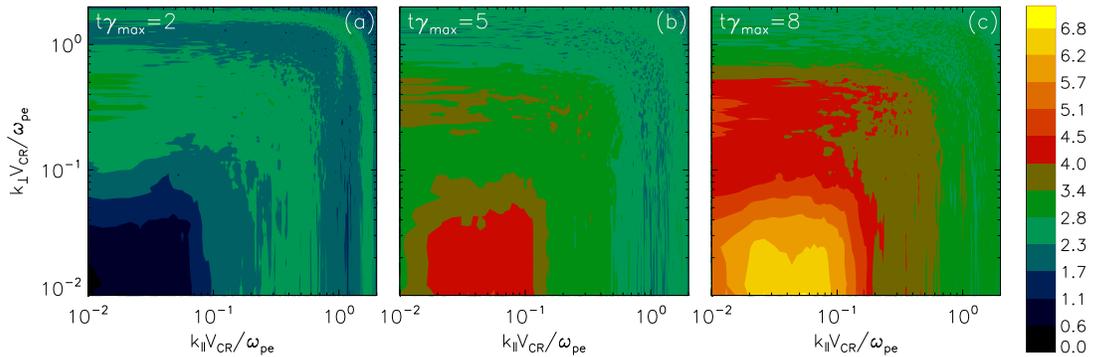}
\caption{Fourier power spectra $[{\rm log}_{10}(dP/d\lambda)]$ of the
perpendicular magnetic-field component $B_z$ in the run with $\gamma_{CR} = 20$
and $v_A = c/20$ (run C) for $t=2,5,$ and $8\gamma_{max}^{-1}$ in
two-dimensional reduced wavevector space $(Z_\parallel, Z_\perp)$. The
wavevector range is chosen to facilitate a direct comparison with analytically
predicted dispersion relation shown in Fig. \ref{fig1}.
\label{four}}
\end{figure}

\begin{figure}
\plotone{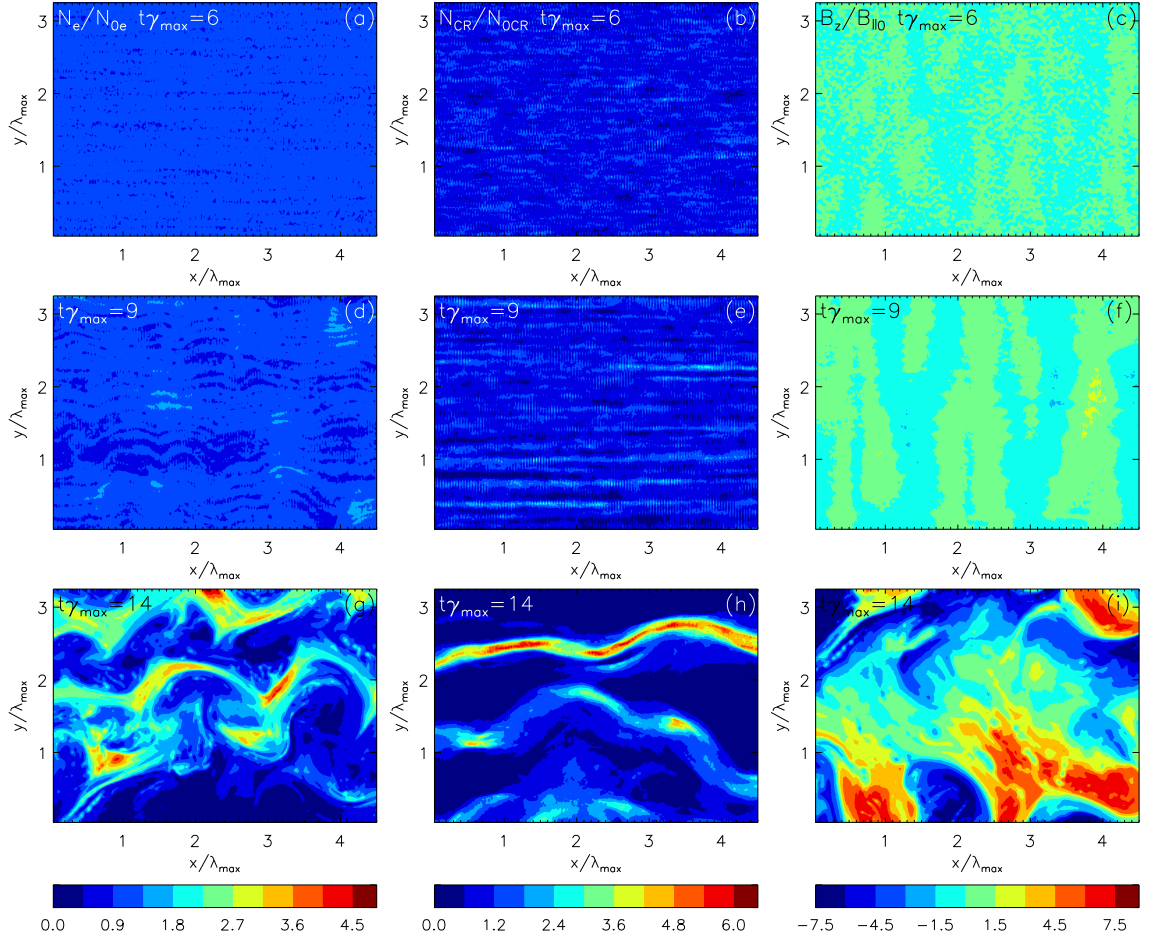}
\caption{Snapshots of the time evolution of the ambient electron density (a, d,
g), cosmic-ray ion density (b, e, h), and the {$B_z$ field amplitude (c, f, i) 
for run C} at 
$t\gamma_{max}=6$ (top), 9 (middle), and 14 (bottom). Electron and ion beam
densities are normalized to their respective initial values. The magnetic field
$B_z$ is normalized to the amplitude of the homogeneous field, $B_{\parallel 0}$.
The spatial scales are provided in units of $\lambda_{max}$ and
only a small portion of the simulation box (($L_{\rm x}, L_{\rm y}) = (10.2
\lambda_{max}, 7.4\lambda_{max})$) is shown.
\label{multi}}
\end{figure}

The characteristic features of magnetic turbulence in a system {\mkp
containing a} mildly
relativistic cosmic-ray beam are detailed in Figures \ref{en16}, \ref{four}, and
\ref{multi} for the run with $\gamma_{CR} = 20$ and $v_A = c/20$ (run C). {The temporal
evolution of the magnetic and electric field average energy densities} is shown in Figure
\ref{en16}. Figure \ref{four} presents Fourier power spectra of the perpendicular
magnetic-field component $B_z$ for $t\gamma_{max}=2,5,$ and $8$ in
two-dimensional reduced wavevector space $(Z_\parallel, Z_\perp)$. Figure
\ref{multi} shows snapshots of the time evolution of the electron and
cosmic-ray ion density, {and the structure in} the $B_z$ magnetic-field component.

The initial filamentation in {the ambient plasma grows quickly in
spatial scale by merging of
adjacent filaments, which can be clearly seen in $E_x$ and 
$E_y$, and also in the $B_z$ field components shown in Figure \ref{four}}.
{\jn Because the Buneman instability between the ion
beam and the plasma electrons is slightly oblique (see Fig. \ref{fig1}c),
hence not purely
electrostatic, it is} visible in magnetic-field Fourier spectra as a
feature at $Z_{\parallel}\approx 1$ (Fig. \ref{four}). The corresponding strong
short-scale modulations in {the densities of ambient electrons and the ion beam} can be
seen in Figure \ref{multi}a-b and Figure \ref{multi}d-e. The nonresonant
parallel modes
of magnetic turbulence emerge in a medium already strongly modified by
filamentation. They appear in a range $0.02\lesssim Z_{\parallel}\lesssim 0.1$
{around the} theoretically predicted $Z_{\parallel}(\lambda_{max})\simeq 0.08$ and
quickly grow in wavelength (see Figs. \ref{four}b-c, \ref{multi}c, and
\ref{multi}f). The influence of the nonresonant modes is stronger on the filamentation
in the slowly drifting ambient plasma than {that} in the relativistic ion
beam. In essence, ambient plasma filaments become vertically tilted (Fig.
\ref{multi}d), which leads to even stronger plasma filamentation. The lack of
spatial correlation between filaments {in the ambient plasma and the beam} 
results in a
local charge imbalance and the build-up of charge-separation electric fields,
which, together with {\jn electric} fields induced by the Buneman instability,
dominate the turbulent electromagnetic energy content of the system in the
initial stage (Fig. \ref{en16}). During the nonlinear stage ($t\gtrsim
8\gamma_{max}^{-1}$) the enhanced filamentation leads to the generation of
{\jn stronger} turbulence in the $B_z$ component of the magnetic field with
perpendicular wavevectors, $k_y$, which disrupts the structure of the parallel magnetic
modes. This interaction between filamentary and nonresonant modes is visible in 
Figure \ref{multi}f and in the Fourier spectrum in Figure \ref{four}c. 

As one can see in Figures \ref{multi}g-i, the strongly amplified magnetic field
starts to backreact on the cosmic-ray beam in the later stage of the system
evolution. Cosmic-ray filaments become tilted and eventually disrupted. This is
accompanied by turbulent ambient plasma motions and results in highly nonlinear
and nearly isotropic magnetic turbulence. 
The characteristics of the turbulence in its post-saturated state are thus
similar to those
observed in {simulations of nonrelativistically drifting hot cosmic-rays}
in the precursor to SNR shocks. 

\subsubsection{Particle Phase-Space Distributions}
\begin{figure}
\plotone{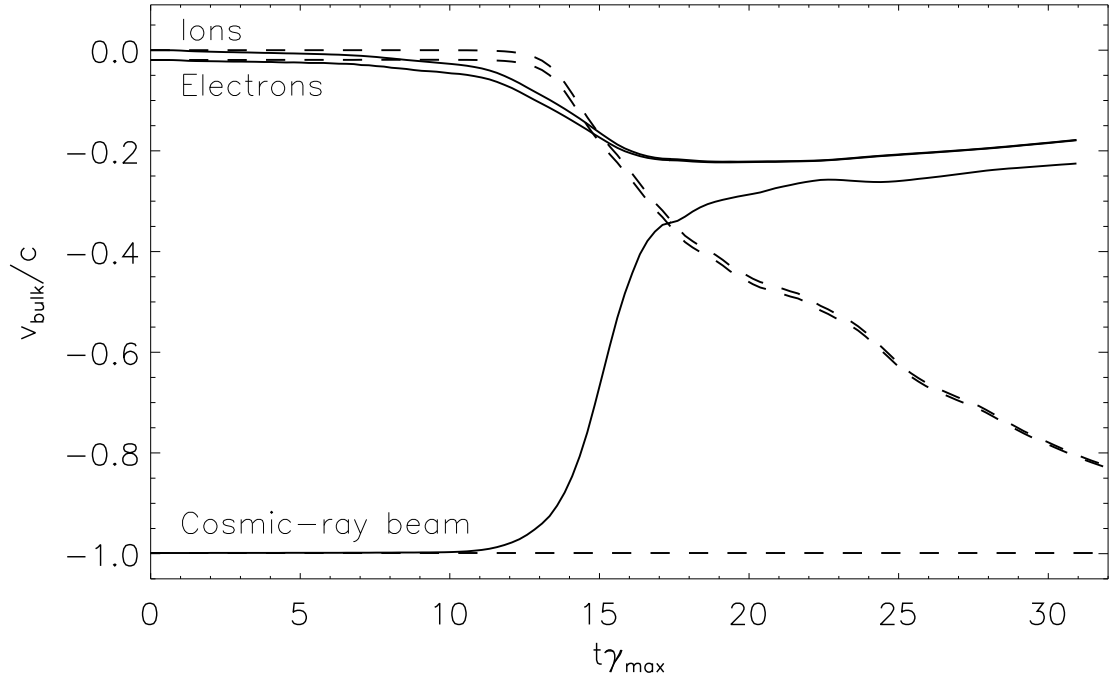}
\caption{Temporal evolution of the bulk (average) velocity of all particle
species
for run A (dashed lines) and run C (solid lines).
\label{vbulk}}
\end{figure}

\begin{figure}
\plotone{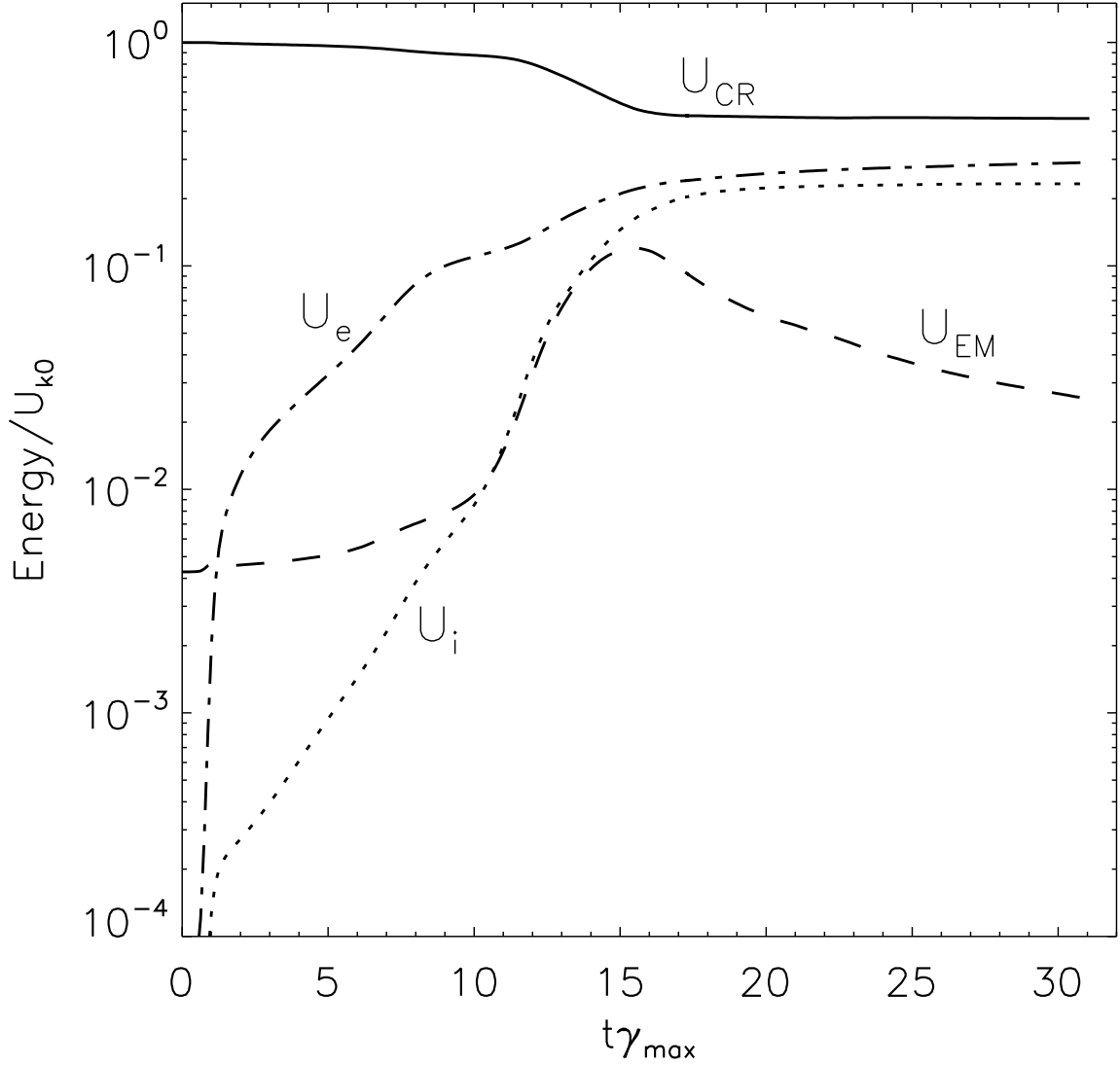}
\caption{Temporal evolution of the average energy density in cosmic-ray beam
ions, $U_{CR}$, ambient electrons, $U_e$, and ions, $U_i$, and the
volume-averaged electromagnetic energy density, 
$U_{EM}=\langle (B_x^2 + B_y^2+B_z^2)/2\mu_0+(E_x^2 + E_y^2+E_z^2)/2\rangle$,
for run C. {\jn All quantities are normalized to the 
initial total kinetic energy in the system, $U_{k0}$.}
\label{energy}}
\end{figure}

\begin{figure}
\plotone{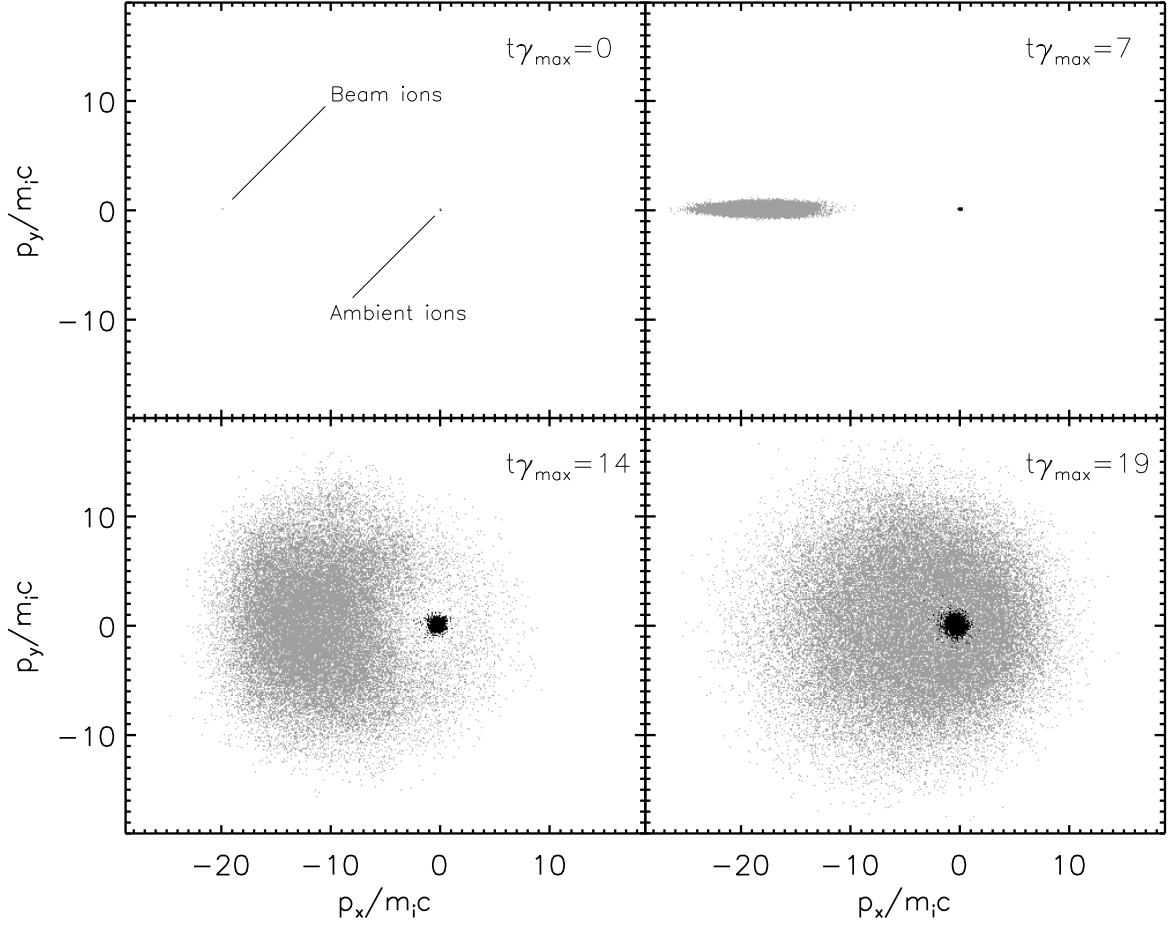}
\caption{{Phase-space distributions of the ions in the cosmic-ray beam (gray) and 
the ambient plasma (black)} at $t\gamma_{max}=0$, 7, 14, and 19 for the run with 
$\gamma_{CR} = 20$ and $v_A =c/20$ (run C). 
\label{phase}}
\end{figure}

The effects of the backreaction of the magnetic turbulence on the particles
are presented in Figures \ref{vbulk}, \ref{energy}, and \ref{phase}. The average (bulk)
velocities of {all} particle species converge in the nonlinear stage, 
{when the magnetic-field growth saturates} (at $t\approx
15\gamma_{max}^{-1}$ for runs A and C in Fig. \ref{vbulk}). While the {
relative drift between the plasma and the cosmic-ray beam disappears in all
our simulations, the mechanism by which that is achieved differs between runs
which suppress the  cosmic-ray backreaction and} fully kinetic runs. 

In the fully kinetic simulations (solid line in Fig. \ref{vbulk}), the cosmic-ray
beam slows down considerably, and the ambient plasma accelerates up to $\sim
-0.2c$. This {behavior is similar} to the results of  \citet{winske} for
nonrelativistic dilute ion beams interacting with ambient plasma via
nonresonant modes, which showed that {the energy of the} decelerating beam
is transferred in
approximately equal parts to the ambient ions and the magnetic field, in
accordance to the predictions of quasi-linear theory. The simulation results of
\citet{winske} were obtained with a one-dimensional hybrid model that treats
electrons as a massless fluid, {with which} one cannot observe
filamentation modes, and the associated electron heating is artificially 
suppressed. As shown in Figure \ref{energy}, which presents the temporal
evolution of energy densities in particles and fields for run C,  in our
simulations the initial plasma filamentation is accompanied by electron
heating at the expense of the beam. However, the heating of the
electrons saturates at $t\approx 7\gamma_{max}^{-1}$, when the nonresonant modes
start to emerge. During the subsequent evolution, beam energy is transferred with
approximately the same rate into the magnetic field and ambient ions, {while the
electrons experience} only moderate further heating. This process of the
energy transfer saturates when the turbulent magnetic field reaches its maximum
energy density and starts to dissipate. The nonlinear evolution of 
{nonresonant modes in a system
containing relativistic ion beams} thus proceeds in
qualitatively the same way as for nonrelativistic beams. 

If the cosmic-ray ion beam is represented by a constant external current, the
energy {\mkp conservation between the beam to the ambient medium is
violated.} As shown in Figure \ref{vbulk} (dashed lines), the saturation of the 
magnetic field growth still comes about by the disappearance of the ion
beam-ambient plasma relative motion, {\mkp at which the return current is provided
without a drift of the plasma electrons relative to the plasma ions,} but now the
ambient ions and
electrons {must assume the constant} cosmic-ray beam bulk velocity.
This {implies that energy is continuously pumped into the
system, and therefore} energy conservation becomes severely violated in the
nonlinear stage. Thus the {validity of simulations that assume a constant 
cosmic-ray current is limited to the early phases in the evolution of the system.}

Figure \ref{phase} shows the {phase-space distributions of the cosmic-ray beam and
the} ambient ions at $t\gamma_{max}=$ 0, 7, 14, and 19 for run C. The early stage of
the system evolution 
($t\gamma_{max}\lesssim 7$)  is dominated by the Buneman instability modes
between the ion beam and ambient electrons.  The electrostatic fields associated
with this mode heat up electrons (Fig. \ref{energy}) and significantly stretch
the beam ions distribution along the beam propagation direction. At the same
time the cosmic-ray beam is {heated} in the transverse direction due to
filamentation modes. The ambient ions remain unaffected by the Buneman
mode\footnote{The phase velocity of the Buneman wave mode between cosmic-ray ion
beam and electrons {\jn is $\sim v_{CR}$ in} the ambient ions rest frame. Thus the
associated electrostatic fields are seen by the ambient ions as a high-frequency
oscillations.}
and become only moderately heated in this stage. The stretching of the beam ions
distribution gradually
saturates at $t\gamma_{max}\sim 7$, by which time the nonresonant modes have set
in and started to strongly backreact on the system.  

During the subsequent evolution the beam momentum becomes quickly randomized in
direction. This randomization is
the combined effect of the pinching of ion-beam filaments and pitch-angle
scattering of the beam particles. The latter process becomes more important in
the highly nonlinear phase ($t\gamma_{max}\gtrsim 10$; compare Fig.
\ref{multi}h), {during which the filaments start to get disrupted.
At the same time the ion beam slows down in bulk, and by 
$t\gamma_{max}\sim 19$ the evolution
saturates when the ion beam particles have been efficiently pitch-angle scattered around
a} mean (bulk) momentum of $\sim -3.8m_ic$. 

The {randomization of beam momentum through} pitch-angle scattering was
previously reported for nonrelativistic beams in conditions allowing for an
efficient magnetic field amplification through nonresonant modes \citep{winske}.
{Here we have demonstrated that these modes can also provide efficient
scattering for relativistic beams.} 

We can estimate the scattering mean free path from the time evolution of the
phase-space distribution of the ion beam, a few snapshots of which are shown in
Figure~\ref{phase}. Between $t\,\gamma_{\rm max}=10$ and $t\,\gamma_{\rm max}=14$
the scattering mean free path in simulation run C is 
\begin{equation}
\lambda_{\rm mfp}\approx 5000\,\Delta. 
\label{mfp}
\end{equation}
At the same time, the rms amplitude of the turbulent magnetic field increases
by more than 250\%, from about $B_{\parallel 0}$ to $3.5\,B_{\parallel 0}$. Using 
the mean of the two numbers, we obtain for the Bohm mean free path 
\begin{equation}
\lambda_{\rm Bohm}\approx 3000\,\Delta 
\label{bohm}
\end{equation}
Given the uncertainty in the estimate arising from the substantial variation in the
magnetic-field amplitude, about a factor 2,
we conclude that the observed scattering mean free path,
and therefore the spatial diffusion coefficient, {\jn for mildly relativistic beams}
are entirely compatible with Bohm diffusion. 

{\jn The estimate of the scattering mean free path can also be made for ultrarelativistic beams based on run B. However,  by the time our simulation ends the CR beam is only partially pitch-angle scattered up to an angle $\sim\pi/6$. Nevertheless, a rough estimate shows that $\lambda_{\rm mfp}$ is again within the factor of a few comparable to $\lambda_{\rm Bohm}$.}

\section{DISCUSSION AND CONCLUSIONS}
We have studied the interaction of a cold, relativistic ion beam penetrating a
cold plasma composed of 
electrons and ions. We have presented 2.5D PIC simulations, complemented with
a linear analysis of the dispersion relation
for linear waves with arbitrary orientation of ${\vec k}$, for parameters that
permit the growth of 
nonresonant, purely-magnetic parallel modes \citep{rev06}. 
Our research is relevant for the understanding of the structure of, and
particle acceleration at, shocks in SNR, GRB, and AGN, for which radiation modeling
suggests that the magnetic 
field near the shock is strongly amplified. 

We observe a close competition of the nonresonant mode with the filamentation
instability and Buneman 
modes, {which is also evident in} the linear dispersion relation. The specific
choice of parameters 
determines which of the three modes of instability dominates. In some cases
filamentation is initially 
important and modifies the later evolution of the parallel nonresonant mode. In
all cases we find that
a representation of the ion beam by a constant current, as is routinely done in
MHD studies, is suboptimal, 
because it suppresses part of the nonlinear response of the system, delays the
saturation processes,
and leads to a significant overestimate of the {final} magnetic-field amplitude. 
 
As in the case of drifting cosmic rays \citep{niem08,stroman} and nonrelativistic
beams \citep{winske},
the saturation of the magnetic-field growth proceeds via bulk acceleration.
For mildly- and {\jn ultra-relativistic} beams, the instability saturates at field
amplitudes {\jn a few times larger than} the homogeneous magnetic field. These results match our recent
studies of nonrelativistically drifting cosmic-rays upstream 
of SNR shocks which also indicated only a moderate magnetic-field amplification
by nonresonant instabilities.

{\jn We have demonstrated that the magnetic field amplified via nonresonant interactions between the CR beam and the plasma can efficiently scatter cosmic rays even for  moderate field amplification levels. 
The scattering mean free path is compatible with Bohm diffusion. Sub-Bohm diffusion was observed in Monte-Carlo simulations of particle transport in the nonlinear turbulent magnetic field generated in the nonresonant instability by \citet{rev08}.  In that work, parallel and perpendicular diffusion coefficients were calculated by probing the spatial displacement of test particles in a static snapshot of the amplified magnetic field that resulted from MHD simulations of the instability. Here we estimate the isotropic spatial diffusion coefficient by probing the evolution of the angular distribution of particles in the self-excited, non-stationary (growing) turbulence whose typical wavelength is at least a factor of a few smaller than the gyroradii of cosmic rays.}  

{In the application to nonrelativistic shocks in SNRs, strong ($\delta B/B_0\gg 1$)
quasi-isotropic magnetic turbulence would be compressed by the shock, thus turning 
into quasi-two-dimensional turbulence in the downstream region. Radio polarimetry suggests that 
the magnetic field immediately behind the shock is preferentially oriented
along the shock normal \citep{sp09}, which is at odds with the above expectation,
if the turbulent field is not quickly damped to an amplitude 
$\lesssim B_{\parallel 0}$ \citep{pyl05}.}

{\jn In the application to relativistic shocks in AGN and GRBs, Monte-Carlo studies of the
first-order Fermi acceleration have shown that the process can operate only for 
quasi-parallel subluminal shocks, provided that strong, short-wave magnetic turbulence exists upstream of the shock \citep{nie06b}. Our results show that the turbulence self-generated by the accelerated particles streaming in the shock precursor may provide scattering sufficient to randomize CR momenta. However,
it is not clear how the strong quasi-isotropic magnetic turbulence in the upstream region influences the particle acceleration at the shock (but see \citet{couch08}). 

Finally, our
simulations show that the saturation of instabilities operating upstream may 
limit the magnetic amplitude to moderate levels. If very strong magnetic field 
is required by radiation modeling, it may therefore be generated at the shock itself or
in the immediate downstream region.}

\acknowledgments
{\jn JN and MP are grateful for the hospitality of Kavli Institute for Theoretical Physics, Santa Barbara, where this work has been completed. JN acknowledges helpful discussions with Mark Dieckmann and Luis Silva.} 
The work of JN is supported
by MNiSW research project N N203 393034, and The Foundation for Polish
Science through the HOMING program, which is supported by a grant from
Iceland, Liechtenstein, and Norway through the EEA Financial
Mechanism.  Simulations were partly performed at the Columbia facility at the
NASA Advanced Supercomputing (NAS). This research was also supported in part by
the National Science Foundation under Grant No. PHY05-51164 and
through TeraGrid resources provided by the
National Center for Supercomputing Applications (NCSA) under project PHY070013N.

\clearpage

\begin{deluxetable}{lccccccc}
\tablecaption{Simulation Parameters and Results}
\tablewidth{0pt}
\tablehead{
Run & Grid & $t^{max}$ & $N_{i}/N_{CR}$ & $\gamma_{CR}$ 
    & $v_{A}$   & $\gamma/\gamma_{max}$ 
    & $\delta B_{\perp}^{max} / B_{\parallel 0}$\\
    & ($\lambda_{max}^2$) & ($\gamma_{max}^{-1}$) &  & & $(c)$ & &}
\startdata
A & 7.4$\times$5.7  & 32.7 & 50  & $\infty$ & 1/20 & 0.83 & 24.0\\
B & 4.6$\times$3.3  & 21.0 & 50  & 300      & 1/20 & 0.94 & 8.9\\
C & 10.2$\times$7.4 & 31.2 & 50  & 20       & 1/20 & 0.43 & 4.7\\
D & 7.4$\times$5.7  & 30.2 & 125 & $\infty$ & 1/50 & 0.79 & 24.0\\
E & 10.2$\times$7.4 & 26.7 & 125 & 20       & 1/50 & 0.4  & 7.5
\enddata
\tablecomments{Parameters and selected results of the simulation runs described
in this paper. Listed are: the system size in units of $\lambda_{max}^2$, the
run duration $t^{max}$ in units of 
$\gamma_{max}^{-1}$, the density ratio of ambient and beam ions, beam Lorentz
factor, the Alfv\'en velocity in units of the speed of light $c$, the measured
growth rate $\gamma$ of the nonresonant modes in units of $\gamma_{max}$, 
and the maximum amplitude of the perpendicular magnetic field perturbations
$\delta B_{\perp}^{max}$ relative to the homogeneous magnetic field. The other
parameters assumed in the simulations are:
the electron skindepth $\lambda_{se}=4\Delta$, the ion skindepth
$\lambda_{si}\simeq 18\Delta$, $\lambda_{max}=562\Delta$, $\gamma_{max}/\Omega_i
= 0.2$, and the ion-electron mass ratio $m_i/m_e=20$.
The runs assuming a constant cosmic-ray current are marked with
$\gamma_{CR}=\infty$. The growth rate $\gamma$ is provided for time periods in
which nonresonant parallel modes show up in the plasma. The amplitudes of the
field turbulence for runs with constant cosmic-ray current (A and D) represent
the unsaturated values at the end of these simulations.}
\end{deluxetable}

\end{document}